\documentclass[12pt]{article}
\usepackage{amsfonts}
 \begin{document}
\begin{center}
\textbf{{\Large Thermal wave packets induced \\by attosecond laser pulses}}

\bigskip
Janina Marciak-Kozlowska\\
Miroslaw Kozlowski

\bigskip
Institute of Electron Technology

Al. Lotnik\'{o}w 32/46, 02-668 Warsaw, Poland

\end{center}

\vspace{2cm}
\begin{abstract}
In this paper the dynamics of the interaction of attosecond laser 
pulses with matter is investigated. It will be shown that the 
master equation: modified Klein-Gordon equation describes the 
propagation of the heatons. Heatons are the thermal wave packets. 
When the duration of the laser pulses 
$\Delta t$
 is of the order of attosecond the heaton-thermal wave packets 
are nondispersive objects. For 
$\Delta t\rightarrow \infty $
 the heatons are damped with damping factor of the order of relaxation 
time for thermal processes.

\end{abstract}

\newpage
\section{Introduction}
The breakthrough progress has been made recently in the generation 
and detection of attosecond laser pulses with high harmonic generation 
techniques. This has heralded an age of attophysics in which 
many electron dynamics will be probed in real time.

In this paper the theoretical framework for attosecond laser 
pulses interaction with matter is developed. In the set of papers 
in \textit{Lasers in Engineering} and in the monograph: \textit{From quarks 
to bulk matter} we formulated the quantum hyperbolic thermal equation 
and find out its solution in different, important from technological 
point of view areas.

The basic notion of the quantum of the thermal field, the heaton 
was introduced in our papers. Up to now the \textit{heatons} wave 
described in the static manner, without the investigation the 
dynamics of the heaton moving. In this paper based on the modified 
Klein-Gordon equation (MK-G) we investigate the motion of free 
heatons. As the result of the study of the solution of MK-G it 
will be shown that the \textit{heatons} are the thermal wave packets. 
For the electron gas the \textit{heatons} represents the packets with 
maximum coincident with the position of the electron in the space.

\section{Physics at the attosecond frontier}
The road from picosecond in femtosecond light pulses has been 
seen laser technology evolve towards lasers that emit light with 
greater spectral width -- that is covering a wide of wavelengths. 
A short pulse results when all the spectral components in light 
beam interfere in a way that adds up to a single burst of light. 
The duration of this pulse is inversely proportional to the spectral 
width. This approach reach its natural limit when the spectral 
span of laser become a significant portion of the visible spectrum. 
This is because the pulse cannot be shorter than a period.

The latest advances are based on research into high-order harmonic 
of femtosecond laser pulses. Precise analysis of the way the 
gas atoms interact with the laser field requires careful application 
of quantum mechanics. The model of P.{\nobreakspace}Corkum{\nobreakspace}[1], 
the original laser pulse tears an electron away from an atom 
and the freed electron moves in response to the laser field being 
accelerated and decelerated as the electromagnetic field oscillates. 
The new harmonics are generated when the electron collides with 
the ion is left behind. This radiation is termed harmonic because 
its frequencies are multiples of the original laser frequency. 
P.{\nobreakspace}Corkum argued that electrons that are ejected precisely 
at the peaks or the crests of the optical pulse are much like 
to radiate. So the radiation is produced in very short burst, 
which occupy just a fraction of the optical cycle. The bursts 
are estimated to last some 100{\nobreakspace}as. In paper{\nobreakspace}[2] 
M.{\nobreakspace}Hentschel et al., start with a 7-femtosecond optical 
pulse and estimate that after filtering more than 90\% of the 
energy of the new radiation they produce is contained in a single 
650 attosecond pulse. 

The work of Hentschel et al.{\nobreakspace}[2] hints at the likely direction 
that attophysics i.e. physics with attosecond laser pulses will 
take in the near future. We have now entered the attosecond world, 
but we will need a better guidebook to help us find the road 
of development.

In monograph{\nobreakspace}[3] the theoretical framework for attosecond 
laser pulse interaction with matter was formulated. It was shown 
that when the laser pulse with duration of the order of attosecond 
interacts with matter it creates the transport processes which 
are described by hyperbolic transport equations.

When the duration of the laser pulse is shorter than the relaxation 
time $\tau $, $\tau =\frac{\hbar }{m\alpha ^{2} c^{2} } $
 ($m =$ electron mass, $c$ is the light velocity and 
$\alpha $  is the fine structure constant, than the transport of thermal 
energy is described by the equation{\nobreakspace}[4]
\begin{equation} 
\frac{1}{v^{2} } \frac{\partial ^{2} T}{\partial t^{2} } +\frac{m_{e}
\gamma }{\hbar } \frac{\partial T}{\partial t} +\frac{2Vm_{e} \gamma
}{\hbar ^{2} } T-\frac{\partial ^{2} T}{\partial x^{2} } =F(x,t).\label{eq1}
\end{equation}
In equation{\nobreakspace}(1) $T$ is temperature, $m_e$ is electron 
mass, $V$ is the potential, 
$\gamma =\left( 1-\alpha ^{2} \right) ^{-\frac{1}{2} } $
 and 
$F(x,t)$
 is the external force.

The solution of equation{\nobreakspace}(1) can be written as
\begin{equation}
T(x,t)=e^{-\frac{t}{2\tau } } u(x,t),\label{eq2}
\end{equation}
where $\tau $  is the relaxation time, $\tau \cong 100\,$as
 i.e. is the order of time duration of attosecond laser pulse. 
After substituting formula{\nobreakspace}(2) to Eq.{\nobreakspace}(1) we obtain
\begin{equation}
\frac{1}{v^{2} } \frac{\partial ^{2} u}{\partial t^{2} } -\frac{\partial
^{2} u}{\partial x^{2} } +qu(x,t)=e^{\frac{t}{2\tau } } F(x,t)\label{eq3}
\end{equation}
and
$$
q=\frac{2Vm}{\hbar ^{2} } -\left( \frac{mv}{2\hbar } \right) ^{2} 
$$
$$m=m_{e} \gamma.$$

In the subsequent we will consider the relativistic electrons, 
without the external forces, i.e. for $F(x,t)\rightarrow 0$. 
In that case Eq.{\nobreakspace}(3) can be written as
\begin{equation}
\frac{1}{v^{2} } \frac{\partial ^{2} u}{\partial t^{2} } -\frac{\partial
^{2} u}{\partial x^{2} } -\left( \frac{mv}{2\hbar } \right) ^{2} u(x,t)=0.\label{eq4}
\end{equation}

\section{Thermal wave packets induced by attosecond laser pulses}
Equation{\nobreakspace}(4) is the modified Klein-Gordon equation which 
can be written as
\begin{equation}
\bar{\Box}u-\left(\frac{mv}{2\hbar}\right)^2u(x,t)=0,\label{eq5}
\end{equation}
where d'Alembert operator $\bar{\Box}$
is equal
\begin{equation}
\bar{\Box}=\frac{1}{v^2}\frac{\partial}{\partial t^2}-\frac{\partial}{\partial x^2}.\label{eq6}
\end{equation}
The ordinary Klein-Gordon equation for the particle with mass $m$ 
is of the form{\nobreakspace}[5]
\begin{equation}
\bar{\Box}u+\left(\frac{m_0}{2\hbar}\right)^2u=0.\label{eq7}
\end{equation}
Equation{\nobreakspace}(5) can be split into its real and imaginary 
parts. Putting for $u(x,t)$
$$
 u(x,t)=\Re (t,x)\exp \left[ \frac{i}{\hbar } S(t,x)\right] 
$$
one obtains
\begin{equation}
\eta ^{ab} (\partial _{a} S)(\partial _{b} S)=\hbar ^{2}
\frac{\bar{\Box}\Re }{\Re}-\left(\frac{mv}{2}\right)^2,\label{eq8}
\end{equation}
where
$$\eta _{ab} ={\rm diag}(1,-1,-1,-1),\quad \quad a,b=1,2,3,4.$$

We use Mackinnon's suggestion{\nobreakspace}[6] therefore we look for 
solutions that satisfy the equation
\begin{equation}
\frac{\bar{\Box}\Re }{\Re}=\left(\frac{mv}{\sqrt{2}\hbar}\right)^2.\label{eq9}
\end{equation}
In this way Eq.{\nobreakspace}(8) becomes
\begin{equation}
\eta ^{ab} (\partial _{a} S)(\partial _{b} S)=m^{2} v^{2} =P_{\mu }
P^{\mu }.\label{eq10}
\end{equation}
If the velocity $v$ is constant, we have from{\nobreakspace}(8)
\begin{equation}
S=-P^{\mu } P_{\mu }\label{eq11}
\end{equation}
and the de Broglie relation 
$P^{v} =\hbar K^{v} $ hold where $K^{v} $
 is the wave number and $P^{v} $  is the classical relativistic four momentum, 
$P^{v} =\left( \frac{E}{v} \vec{p} \right) $.
In paper{\nobreakspace}[6] L.{\nobreakspace}Mackinnon constructs a wave packet 
considering that a wave 
$\Phi =\Phi _{0} \exp \left[ i\omega t\right] $
 of frequency 
$\omega =\frac{mc^{2} }{\hbar } $
 is associated with a particle of rest mass $m$ and that for 
an observer moving with a constant velocity 
$v$
 with respect to the particle, the associated wave (be means 
of a Lorentz transformation) acquires the form 
$\Phi =\Phi _{0} \exp \left[ i\omega \gamma \left( t-\frac{v}{c^{2} }
x\right) \right] $.
Mackinnon showed the wave packet is compatible with the basic 
experiments of quantum mechanics and does not spread in time 
the found
\begin{equation} 
\Re =A\frac{\sin[gr]}{gr},\label{eq12}
\end{equation}
where $A=$ constant and 
$g=\frac{mv}{\sqrt{2} \hbar } $, and 
$g=\frac{m\alpha c}{\sqrt{2} \hbar}  $
 and
\begin{equation}
r=\gamma (x-vt)\label{eq13}
\end{equation}
is the distance from the particle portion, so that
\begin{equation}
u(x,t)=A\frac{\sin gr}{gr} \exp \left[ -\frac{i}{\hbar } (Et-px)\right] \label{eq14}
\end{equation}
is the Lorentz boost of the solution
$$\Phi =A\frac{\sin gr'}{gr'} \exp \left[ -i\omega t'\right],$$
where $r'=x$. This solution was first found by de Broglie{\nobreakspace}[7] and 
also used in the stochastic interpretation of quantum mechanics 
by Vigier and Gueret{\nobreakspace}[8].

In the monograph{\nobreakspace}[3] the program of the quantization of 
the temperature field was undertaken. The master quantum equation 
for the temperature field was equation{\nobreakspace}(5). The quantum 
of temperature field, the \textit{heaton} was introduced. The \textit{heaton} energy 
$E=\hbar \omega $, with 
$\omega =\frac{mv^{2} }{\hbar } $, 
$\omega =\frac{m\alpha ^{2} c^{2} }{\hbar } $. Considering 
equation{\nobreakspace}(5) the \textit{heaton} is the thermal 
wave packet which maximum coincides with electron and which shape 
is described by equation{\nobreakspace}(14).
\section{Conclusions}
In this paper the modified Klein-Gordon (MK-G) equation for temperature 
field was obtained. It was shown that the solution of MK-G equation 
is the \textit{quantum} of the temperature field, \textit{heaton} (introduced 
in our paper \textit{Found. of Physics Letters}, \textbf{9}, (1996) p.{\nobreakspace}235). 
The \textit{heaton} is the thermal wave packet which maximum coincides 
with electron and which is moving with velocity{\nobreakspace}
$v =\alpha c$. As the result of the interaction of the attosecond laser pulses 
with electron gas the \textit{heatons} are created. When the laser 
pulse is shorter than the relaxation time 
$\tau =\frac{\hbar }{mv^{2} } $
 the \textit{heaton} are not dispersed.

\end{document}